# Enhanced Cherenkov radiation in twisted hyperbolic Van der Waals crystals

Hao Hu[1,*], Xiao Lin[2,3], Guangwei Hu[4], Francisco J. Garcia-Vidal[5], and Yu Luo[1,*]

[1]*National Key Laboratory of Microwave Photonics, Nanjing University of Aeronautics and Astronautics, Nanjing 211106, China*

[2]*Interdisciplinary Center for Quantum Information, State Key Laboratory of Modern Optical Instrumentation, Zhejiang University, Hangzhou 310027, China.*

[3]*International Joint Innovation Center, Key Lab. of Advanced Micro/Nano Electronic Devices & Smart Systems of Zhejiang, The Electromagnetics Academy at Zhejiang University, Zhejiang University, Haining 314400, China.*

[4]*School of Electrical and Electronic Engineering, Nanyang Technological University, 50 Nanyang Avenue, Singapore 639798, Singapore*

[5]*Departamento de Física Teórica de la Materia Condensada and Condensed Matter Physics Center (IFIMAC), Universidad Autónoma de Madrid, Madrid E-28049, Spain*

[*]*Corresponding authors. Email: hao.hu@nuaa.edu.cn (H. Hu); yu.luo@nuaa.edu.cn (Y. Luo)*

**Cherenkov radiation in artificial structures experiencing strong radiation enhancements promises important applications in free-electron quantum emitters, broadband light sources, miniaturized particle detectors, etc. However, the momentum matching condition between the swift electron and emitted photons generally restricts the radiation enhancement to a particular momentum. Efficient Cherenkov radiation over a wide range of momenta is highly demanded for many applications but has still remained a challenging task. To this end, we explore the interaction between a swift electron and twisted hyperbolic Van der Waals crystals, and observe enhanced Cherenkov radiation at the flatband resonance frequency. We show that, at the photonic magic angle of the twisted crystals, the electron momentum, once matching with that of the flatband photon, gives rise to a maximum energy loss (corresponding to the surface phonon generation), one-order of magnitude higher than that in conventional hyperbolic materials. Such a significant enhancement is attributed to the excitation of flatband surface phonon polaritons over a broad momentum range. Our findings provide a feasible route to highly directional free-electron radiation and radiation shaping.**

**Introduction**

Cherenkov radiation refers to a directional photon emission from a swift charged particle. To enable Cherenkov radiation, the swift charged particles should fulfill the Cherenkov phase-matching condition, i.e., $\bar{k} \cdot \bar{v}_e = \omega$ [1]. Here, $\bar{k}$ and $\omega$ correspond to the wavevector and frequency of emitted photonic modes, respectively; $\bar{v}_e$ accounts for to the particle velocity. Particularly, in a homogenous, dispersionless and isotropic medium, this Cherenkov phase-matching condition could be simplified as $\cos\theta_c = 1/(n\beta)$ [2], where $\theta_c$ is the angle between the photon emission direction and particle trajectory, $n$ is the refractive index of the medium, $\beta = v_e/c$ is the normalized velocity, and $c$ is the light speed in vacuum. Such a unique relation is widely applied to detect particle masses by measuring the Cherenkov angles at a given particle momentum[3], leading to the famous discovery of elementary particles such as anti-proton and J-particle[4,5].

Cherenkov phase-matching condition also indicates the flexibility to engineer the behaviors of Cherenkov radiation by controlling the material and structural dispersions [6,7]. For example, double negative metamaterials, whose refractive indices are negative, have been proposed to reverse the direction of Cherenkov radiation [8-10]. A swift electron can emit surface Dyakonov-Cherenkov radiation only at a particular particle trajectory, owing to the narrow angular existing domain of Dyakonov surface waves [11]. By exploiting the high-wavevector (high-k) modes in plasmonic structures, one can largely reduce the velocity threshold of Cherenkov radiation and enhance the photon emission rate at a given frequency [12-15]. Recent advances further demonstrate the broadband enhancement of Cherenkov radiation by interacting free electrons with the dispersionless surface plasmons [16]. These novel effects broaden the applications of Cherenkov radiation in the miniaturized particle detectors[3,17,18], novel X-ray/terahertz light sources [19-24], molecular imaging [25], and radiation therapy[26], among others.

Despite great effort made to enhance Cherenkov radiation, the free electron generally emits the photonic modes with a particular wavevector $|\bar{k}| = \omega/(v_e \cos\theta_p)$ at a particular frequency. Here, $\theta_p$ refers to the angle between the particle trajectory and the wavevector of emitted photonic modes, and $\theta_p = \theta_c$ only in homogeneous isotropic media. This constraint limits the maximum achievable emission rate of swift electrons. To break this limit, the Cherenkov phase-matching condition has to be satisfied in a wide range of $\theta_p$, and hence $|\bar{k}|$. A recent experimental work has demonstrated a feasible solution to enhance free-electron radiation in a finite range of $|\bar{k}|$ by exploiting the flatband Bloch modes in all-dielectric photonic crystals [27]. However, the mode indices of Bloch modes in all-dielectric photonic crystals are close to the unity. This not only makes the existing domain of flatband modes relatively narrow in the momentum space, but also necessitates the electron velocity comparable to the light speed in vacuum. Thus, a novel scheme to enhance Cherenkov radiation in a broad range of wavevectors is highly desired for constructing, e.g., high-performance free electron light sources on chip.

To this end, we propose flatband surface phonon polaritons to enhance Cherenkov radiation in a broad range of wavevectors. The flatband surface phonon polaritons widely exist in the twisted hyperbolic Van der Waals crystals [28-31]. As the photonic analogue of moire superlattice in condensed matter physics [32,33], the twisted hyperbolic Van der Waals crystals can enable flatband surface phonon polaritons with the flat isofrequency contour at a relatively large twisted angle (i.e., the photonic magic angle[29]). Remarkably, such surface modes are featured with large mode indices (generally over 100). Thus, flatband surface phonon polaritons open up a new opportunity to enhance Cherenkov radiation in a broad range of wavevectors with low-energy electrons. Here, we investigate the interaction between the swift electron and the twisted hyperbolic Van der Waals crystal α-MoO$_3$ that supports flatband surface phonon polaritons. With the suitable selection of electron velocity, Cherenkov radiation is mediated by flatband surface phonon polaritons. Our studies show that the wavevector linewidth of photon emission from the swift electron is enhanced by nearly 20 times at the photonic magic angle, as that compared to the conventional ones. As such, the maximum achievable photon number of Cherenkov radiation is one-order of magnitude higher than that in an individual α-MoO$_3$ slab. Furthermore, the particle velocity that leads to the maximum achievable photon number could be readily controlled by tuning the gap width of the twisted α-MoO$_3$ slabs. Our work not only facilitates the realization of free-electron light sources on chip, but also could find applications in the particle detection.

**Results**

Without loss of generality, we consider a twisted hyperbolic Van der Waals material composed of two stacked α-MoO$_3$ slabs. The permittivity of α-MoO$_3$ is adopted from the recent work [34] (as detailed in Section S1 of supplementary material). The thickness of each α-MoO$_3$ slab is denoted as $h$, and the gap width between two α-MoO$_3$ slabs is denoted as $g$. The twisted angle of the top- (bottom-) layer α-MoO$_3$ slab with respect to x-axis is $\phi/2$ ($-\phi/2$). As such, the twisted angle between two α-MoO$_3$ slabs is $\phi$. A free electron (with the current density of $\bar{J} = \hat{x} q_0 \delta(y) \delta(z) \delta(x - v_e t)$) is moving along x-axis at the velocity of $v_e$, where $q_0$ is the elementary charge [35]. For conceptual demonstration, the free electron is travelling in the gap of twisted hyperbolic material.

We first demonstrate that the twisted angle of α-MoO$_3$ slabs can strongly influence Cherenkov radiation. To illustrate this, we plot in Fig. 1(b&c), the photon number as a function of the wavevectors $k_y = \bar{q}/\sin\theta_p$ in different twisted angles, when the studied wavenumber is fixed at $\nu = 900$ cm$^{-1}$. Here, $\bar{q}$ refers to the in-plane wavevector of photon polaritons. As shown in Fig. 1(b), the free electron can emit the photons in a broad range of wavevectors (with the normalized linewidth up to 21.3) when the twisted angle equals to the photonic magic angle, i.e., $\phi = \phi_{magic} = 44.07°$. In sharp contrast, the free electron only emits photons in a narrow range of wavevectors (with the normalized linewidth of 1.2), if $\phi \neq \phi_{magic}$ [Fig. 1(c)]. Here, the photonic magic angle $\phi_{magic}$ is known as the one that can enable the open-to-close topological transition of isofrequency contour of surface phonon polaritons, accompanied with the variation of number of anti-crossing points (the anti-crossing point is defined as the crossing point of isofrequency contours when the constituent materials are uncoupled) [28]. The magic angle is theoretically calculated by $\phi = 2\varphi$ [28,29], with $\varphi$ to be the open angle of isofrequency contour of an individual α-MoO$_3$ slab [see black solid line in the inset of Fig. 2(c)]. Specifically, at $\nu = 900$ cm$^{-1}$, the magic angle is $\phi_{magic} = 44.07°$ where open-to-close topological transition emerges: if $\phi = 20° < \phi_{magic}$ at $\nu = 900$ cm$^{-1}$, the isofrequency contour is an open hyperbola with 2 anti-crossing points; if $\phi = 60° > \phi_{magic}$ at $\nu = 900$ cm$^{-1}$, the isofrequency contour becomes an open hyperbola with 4 anti-crossing points [seeing the inset of Fig. 2(c)].

The broad wavevector spectrum at the photonic magic angle largely enhances Cherenkov radiation. To demonstrate this, we plot in Fig. 2(a) the photon number as a function of twisted angle and inverse of the normalized velocity, where the inverse of the normalized velocity is related to the realistic electron velocity by $1/\beta = c/v_e$, when $\nu = 900$ cm$^{-1}$. Obviously, the photon number reaches a maximum value, i.e., $3.1\times10^{-8}$ (m·rad/s)$^{-1}$, only when $\phi = 43.35°$ and $1/\beta = 10.78°$ [see Fig. 2(a)]. By contrast, the maximum photon numbers are $0.8\times10^{-8}$ and $2.1\times10^{-8}$, when the twisted angles are $\phi = 35°$ and $\phi = 50°$, respectively [see Fig. 2(b)]. The twisted angles that enable the maximum photon number are highly close to the photonic magic angles in theory, indicating that the broad wavevector spectrum at the photonic magic angle plays a key role to enhance Cherenkov radiation [see Fig. 2(c)].

To further explain the reason for the emergence of broad wavevector spectrum, we analyze the field patterns of free-electron radiation in the twisted α-MoO$_3$ slabs. Fig. 3(a-c) shows the field pattern of Cherenkov radiation, when we vary the twisted angle $\phi$ from $35°$, $43.35°$ to $50°$ and the inverse of the normalized velocity is fixed at $1/\beta = 10.78$. When $\phi = 35°$, the isofrequency contour of surface photon polaritons is hyperbolic. As such, a free electron with $1/\beta = 10.78$ only emits two photonic modes satisfying $\bar{q} \cdot \bar{v}_e = \omega$ [see Fourier spectrum of Cherenkov radiation in Fig. 3(d)]. When $\phi = 43.35°$, the isofrequency contour becomes nearly flat. In this case, modes in a broad range of wavevectors, i.e., from $k_y = 40k_0$ to $k_y = 60k_0$ satisfy the Cherenkov phase-matching conditions. These modes thus are efficiently excited by the free electrons [see Fourier spectrum of Cherenkov radiation in Fig. 3(e)]. With the further increase of twisted angle, e.g., $\phi = 50°$, no surface mode is excited by the free electron [see Fourier spectrum of Cherenkov radiation in Fig. 3(f)].

To quantitively demonstrate the influence of the twisted angle on the wavevector spectrum, we plot the photon number as a function of $k_y$ and $\phi$ in Fig. 3(g). We extract linewidth $\Delta k_y$ of Cherenkov radiation from the Fig. 3(g). The twisted angle is varied from $20°$ to $43.35°$, the linewidth $\Delta k_y$ is enhanced by 13 times (if the material loss rate is $\gamma = 4$ cm$^{-1}$, which is the typical value for α-MoO$_3$). Such an enhancement rate can even reach 43 when the material loss rate is reduced from 4 cm$^{-1}$ to 1 cm$^{-1}$.

Finally, we investigate the influence of the gap width $g$ on the behaviors of Cherenkov radiation mediated by flatband surface phonon polaritons. Fig. 4 plots the photon number as a function of normalized velocity and gap width of α-MoO$_3$ slabs, where the twisted angle is fixed at the photonic magic angle. As the gap width increases, the electron velocity that enables the maximum photon number increases [Fig. 4(a)]. To be specific, the normalized electron velocities $\beta_{\text{peak}}$ for the maximum achievable photon number are 0.075, 0.093, and 0.106, respectively, when the gap widths are 2 nm, 5 nm and 10 nm, respectively [Fig. 4(b)]. This is because the gap width determines the mode confinement of flatband surface photon-polariton. As the gap width decreases, the coupling strength of photon-polariton mode in each flake becomes stronger, introducing larger dispersion line splitting in momentum space and thus pushing the $k_x$ (of the mode closest to the origin) larger. That is, to excite such flatband surface photon-polaritons with a larger $k_x$, a smaller electron velocity is required. Such an effect may inspire novel particle detectors based on enhanced Cherenkov radiation in twisted hyperbolic Van der Waals crystals.

**Discussions**

To conclude, we report on Cherenkov radiation mediated by flatband surface phonon polaritons, in twisted hyperbolic Van der Waals crystals at the photonic magic angle. Our revealed Cherenkov radiation comprises of photons in a broad range of wavevectors at the working frequency. Such behavior largely enhances emission rate of a free electron in passive media, leading to the maximum electron energy loss at the working frequency. Via altering the twisted angle between two hyperbolic materials, the flatband resonance frequency could be flexibly tuned from the terahertz to mid-infrared frequency range. Our revealed Cherenkov radiation thus finds applications in the high-intensity tunable free-electron light sources applied to the terahertz/mid-infrared frequency. On the other hand, since the excitation of enhanced Cherenkov radiation highly relies on the twisted angle, our technique offers a potential solution to measure the photonic magic angle in twisted photonic structures. Moreover, as the excitation of enhanced Cherenkov radiation is also highly sensitive to the particle velocity, our system may open a new door for particle selection. We remark that our mechanism could be extended to multilayered twisted photonic systems, opening up an opportunity to enhance Cherenkov radiation in 3D scenarios [36]. Finally, we highlight that our findings are general as it can apply to other polaritonic system and a broad range of hyperbolic materials [37,38], including hyperbolic metasurface [28], hexagonal boron nitride [39], vanadium pentoxide [40], dichalcogenide tungsten diselenide [41], black phosphorus [42] or their hybrids [43].

**Author Contribution**
H.H. initiated the project. Y.L. supervised the project. H.H. performed the analytical calculation and numerical simulations. All authors contributed to the analysis and discussion of the results. H.H. and Y.L. wrote the paper. All authors have read and approved the final version of the manuscript.

**Declaration of competing interest**
The authors declare that they have no known competing financial interests or personal relationships that could have appeared to influence the work reported in this paper.

**Data availability**
Data will be made available on request.

**Acknowledgement**
This work is sponsored in part by Distinguished Professor Fund of Jiangsu Province (Grant No. 1004-YQR23064, 1004-YQR24010); Selected Chinese Government Talent-recruitment Programs of Nanjing (Grant No. 1004-YQR23122); Startup Grant of Nanjing University of Aeronautics and Astronautics (Grant No. 1004-YQR23031); National Natural Science Fund for Excellent Young Scientists Fund Program (Overseas) of China; National Natural Science Foundation of China (Grant No. 62175212); Zhejiang Provincial Natural Science Fund Key Project (Grant No. LZ23F050003); the Fundamental Research Funds for the Central Universities (Grant No. 226-2024-00022).

**Appendix A. Supplementary data**

Supplementary data to this article can be found online at https://doi.org/XXX.

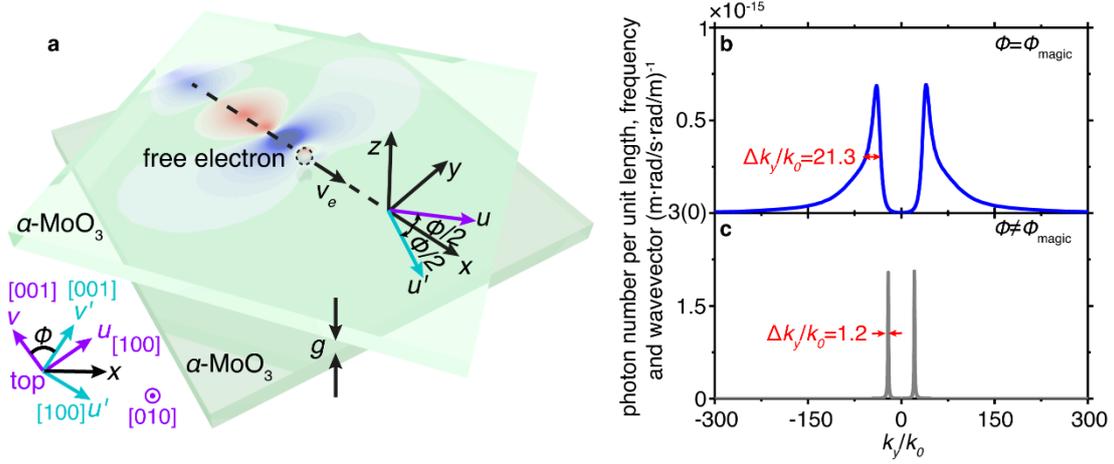

**Figure 1. Schematic of Cherenkov radiation in twisted hyperbolic Van der Waals crystals.** (a) Structural setup. Two α-MoO₃ slabs are stacked in the z-direction. The u-axis and v-axis are oriented in the [100] and [001] directions for the top-layer α-MoO₃ slab, respectively; The u′-axis and v′-axis are oriented in the [100] and [001] directions for the bottom-layer α-MoO₃ slab, respectively. The twisted angle of the top- (bottom-) layer α-MoO₃ slab with respect to x-axis is $\phi/2$ ($-\phi/2$). As such, the twisted angle between two α-MoO₃ slabs is $\phi$. A free electron with the velocity of $v_e$ propagates inside the gap of twisted α-MoO₃ slabs along x-axis. (b) Photon number spectrum at the magic angle $\phi = \phi_{magic}$. (c) Photon number spectrum at the conventional angle $\phi \neq \phi_{magic}$. In (b,c), $k_y$ corresponds to the tangential wavevector of photons emitted by the free electron. Without particular statement, other parameter setups here and below are: the inverse of the normalized velocity of free electron is $1/\beta = 10.78$; the thickness of each α-MoO₃ slab is $h = 100$ nm; the gap width between two α-MoO₃ slabs is $g = 5$ nm; the studied wavenumber is $\upsilon = 900$ cm⁻¹; the magic angle is $\phi_{magic} = 44.07°$ at $\upsilon = 900$ cm⁻¹. Moreover, optical parameters of each α-MoO₃ are provided in the section S1 of supplementary material.

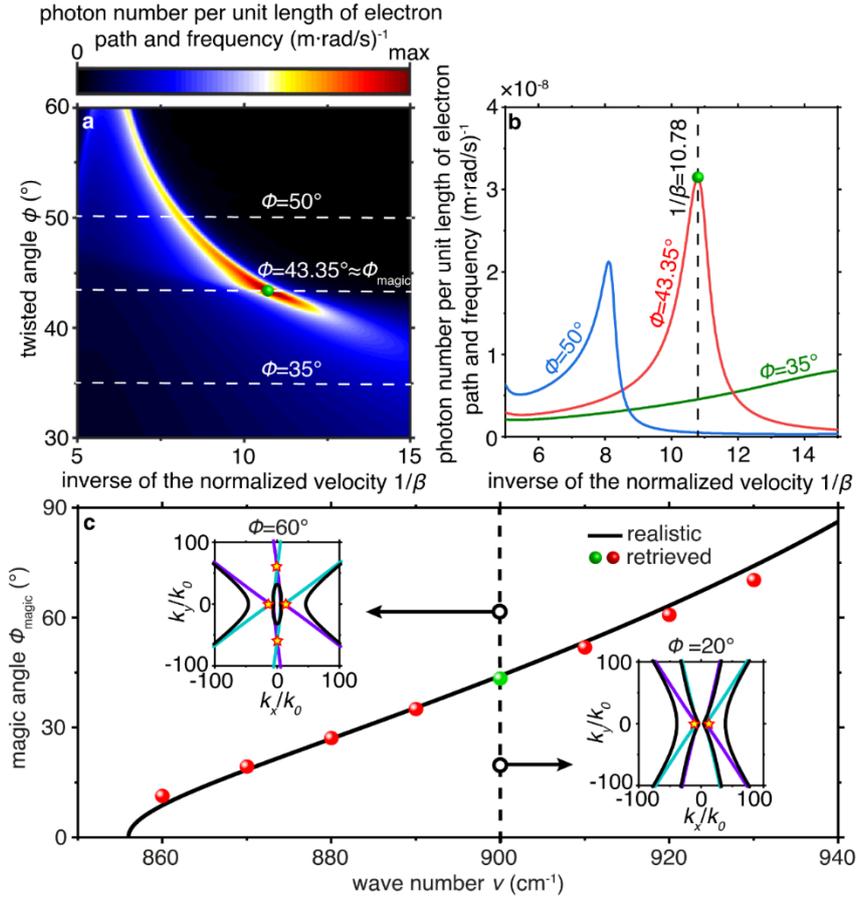

**Figure 2. Enhancing Cherenkov radiation by flatband surface photon polaritons.** (a) Photon number as a function of twisted angle and inverse of the normalized velocity, defined as $1/\beta = c/v_e$. (b) Photon number as a function of inverse of the normalized velocity. The studied twisted angles are $\phi = 35°$, $\phi = 43.35°$ and $\phi = 50°$, respectively. (c) Comparison of retrieved and realistic magic angles. The realistic magic angle corresponds to the one that enables the open-to-close topological transition of isofrequency contour of surface phonon polaritons [see the isofrequency contours before and after the topological transition at $\upsilon = 900$ cm$^{-1}$ in the inset of Fig. 2(c)]. The retrieved magic angle corresponds to the one allowing the maximum photon number in Fig. 2(a). In each inset, the blue and purple curves correspond to the isofrequency contours of uncoupled surface phonon polaritons; the black curve corresponds to the isofrequecy contour of coupled surface phonon polaritons.

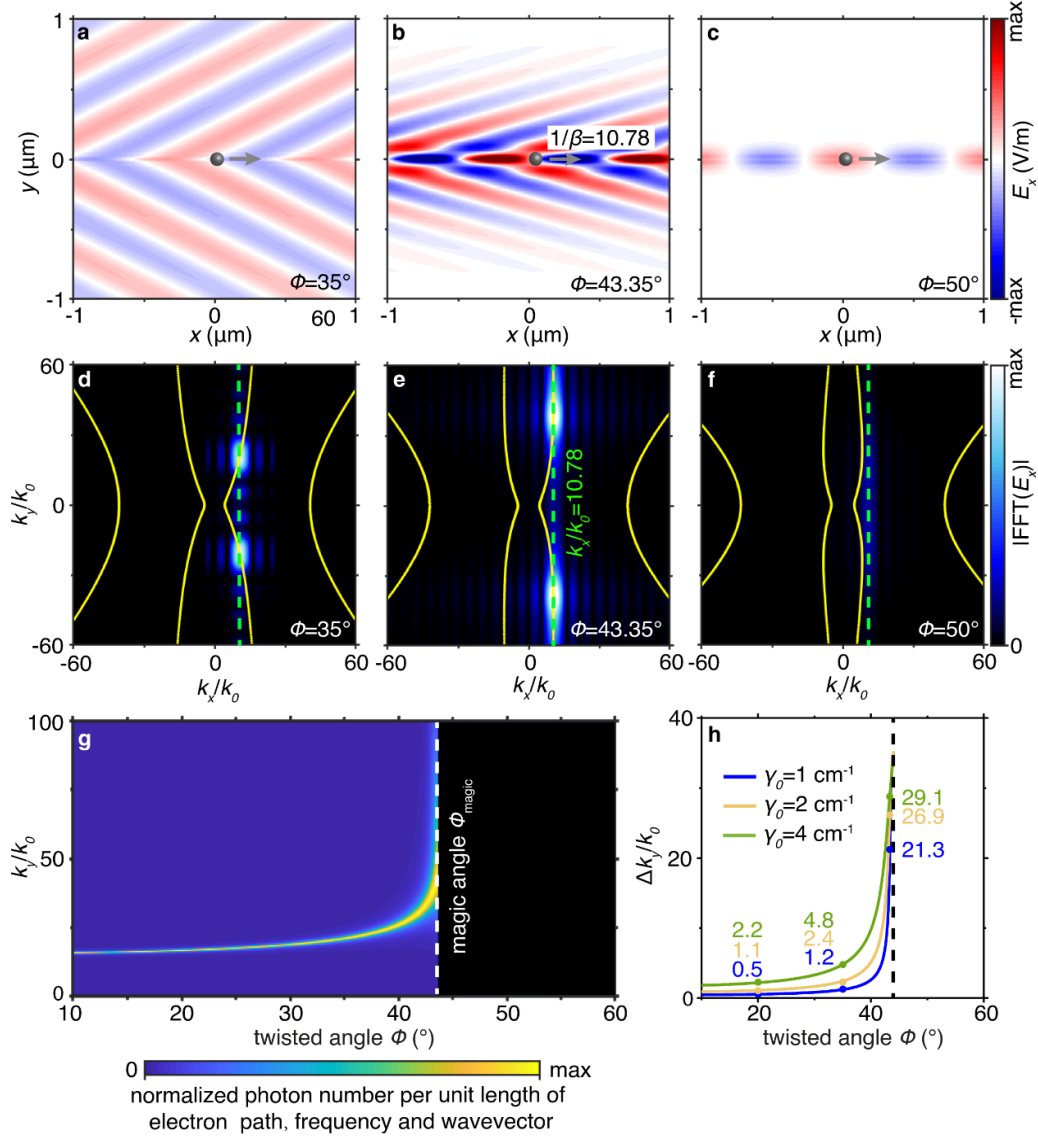

**Figure 3. Field patterns and its Fourier spectrum of Cherenkov radiation mediated by flatband and hyperbolic surface phonon polaritons.** (a,c) Field patterns of Cherenkov radiation mediated by surface phonon polaritons. The twisted angles of α-MoO$_3$ slabs are $\phi = 35°$, $\phi = 43.35°$ and $\phi = 50°$, respectively. (d,f) Fourier spectrum of field patterns in Fig. 3(a-c). The yellow solid curves correspond to the theoretical isofrequency contours of surface photon polaritons. (g) The normalized photon number emitted by free electron at $k_x/k_0 = 10.78$ as a function of twisted angle. (f) The linewidth of $k_y$ as a function of twisted angle at $k_x/k_0 = 10.78$. The studied loss rates are $\gamma = 1$ cm$^{-1}$, $\gamma = 2$ cm$^{-1}$ and $\gamma = 4$ cm$^{-1}$, respectively.

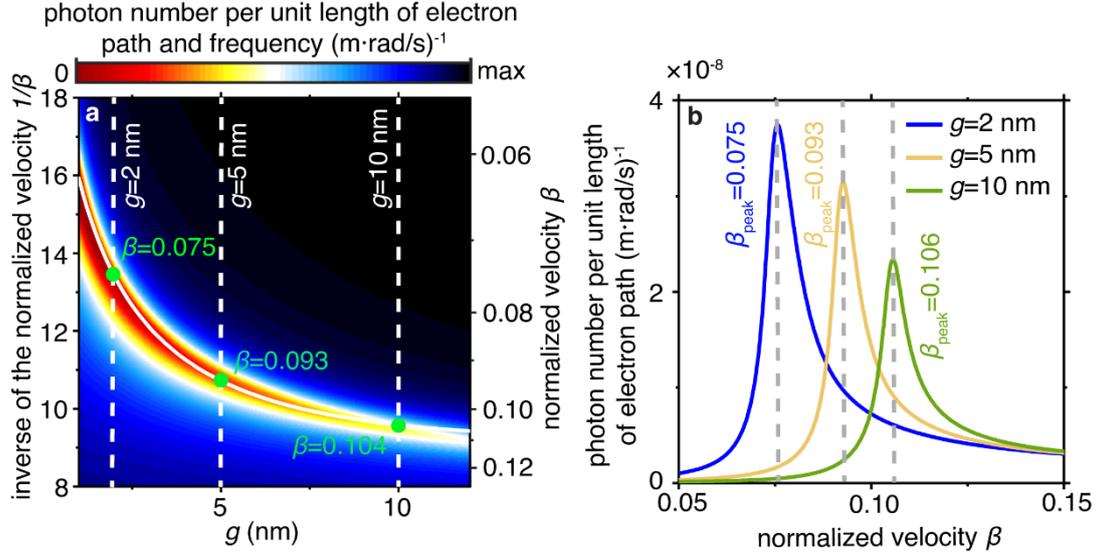

**Figure 4. Influence of the gap width on the behaviors of Cherenkov radiation mediated by flatband surface phonon polaritons.** (a) Photon number as a function of gap width and inverse of the normalized velocity. (b) Photon number as a function of normalized velocity. The studied gap widths are $g = 2$ nm, $g = 5$ nm, and $g = 10$ nm, respectively. In (a), the white-solid curve corresponds to the theoretical value of normalized velocity to excite flatband surface phonon polaritons at different gap width of twisted α-MoO$_3$ slabs. In (b), the $\beta_{peak}$ corresponds to the normalized velocity that enables the maximum photon number.